\documentclass[epj]{svjour}
\usepackage{graphicx}
\usepackage{epsfig}
\usepackage{dcolumn}
\usepackage{amsmath}
\usepackage{bm}
\usepackage{array}
\usepackage{hyperref}

\newcommand{\PreserveBackslash}[1]{\let\temp=\\#1\let\\=\temp}

\newcolumntype{C}[1]{>{\PreserveBackslash\centering}p{#1}}
\newcolumntype{R}[1]{>{\PreserveBackslash\raggedleft}p{#1}}
\newcolumntype{L}[1]{>{\PreserveBackslash\raggedright}p{#1}}
\makeatletter
\newcommand{\figcaption}{\def\@captype{figure}\caption}

\newcommand{\tabcaption}{\def\@captype{table}\caption}
\makeatother
\begin{document}
\title{$B_c$ meson rare decays in the light-cone quark model}
\author{\textbf{Teng Wang}\inst{1} \and \textbf{Tianbo Liu}\inst{1} \and \textbf{Da-Xin Zhang}\inst{1} \and  \textbf{Bo-Qiang
Ma}\inst{1}$^,$\inst{2}$^,$ \thanks{e-mail: mabq@pku.edu.cn}
}                     
\institute{School of Physics and State Key Laboratory of Nuclear
Physics and Technology, Peking University, Beijing 100871, China
\and Center for High Energy Physics, Peking University, Beijing
100871, China}
\date{Received: date / Revised version: date}
%
\abstract{ We investigate the rare decays $B_c \rightarrow D_s(1968)
\ell \overline{\ell}$ and $B_c\rightarrow D_s^*(2317) \ell
\overline{\ell}$ in the framework of the light-cone quark
model~(LCQM). The transition form factors are calculated in the
space-like region and then analytically continued to the time-like
region via exponential parametrization. The branching ratios and
longitudinal lepton polarization asymmetries~(LPAs) for the two
decays are given and compared with each other. The results are
helpful to investigating the structure of $B_c$ meson and to testing
the unitarity of CKM quark mixing matrix. All these results can be
tested in the future experiments at the LHC.
%
} 
\maketitle
\section{Introduction}

The investigation on heavy-quark mesons is an active 
frontier of particle physics. The study of heavy-quark meson decays
not only gives us insights on the hadron structure such as the
hadron wave function~\cite{Huang:2004na} and the hadron transverse
momentum distribution~\cite{Ma,Li2004ay}, but also provides us an
ideal field to study the mixing between different generations of
quarks by extracting the Cabibbo-Kobayashi-Maskawa~(CKM) matrix
elements. The investigation of 
the CKM matrix elements in heavy-quark meson decay processes can
help us to test the charge-parity~(CP) violation in the standard
model~(SM)~\cite{He:2000ys,He:2004wha,He:2004whb,He:2004whc,He:2004whd}
and to search for new
physics beyond 
the SM~\cite{Du:1997pma,Du:1997pmb}. Among all the heavy-quark
mesons, the $B_c$
meson is of special interest 
because of its some unique properties. It is the lowest bound state
composed of two heavy quarks ($b$ and $c$) with explicit flavor
numbers. Distinguished from other heavy quark bound states like
charmonia ($c\bar{c}$ bound state) and bottomonia ($b\bar{b}$ bound
state) with implicit flavor numbers, $B_c$ can only decay via weak
interaction. Thus, $B_c$ meson provides us a chance to study the
weak interaction and the CKM matrix elements with all three
generations included. Compared with the study of the $B$ meson, the
$B_c$ meson received less attention, because the production of $B_c$
mesons requires a much higher energy which is unaccessible to most
available colliders. However, it was predicted that $B_c$ mesons can
be generated dramatically via different
ways~\cite{Chang1a,Chang1b,Chang1c,Chang2a,Chang2b,Chang2c,Chang2d}
in experiments at the Large Hadron Collider~(LHC) which is running
now. Therefore, it is mature for us to study the $B_c$ meson on many
of its physical quantities experimentally.

Among all the $B_c$ meson decay modes, the rare decays $B_c
\rightarrow D_s(1968,2317)\ell \overline{\ell}$ induced by
the flavor changing neutral currents~(FCNCs) are the most exciting
ones. FCNCs processes have received lots of attention since the
CLEO's measurement of the radiative decay
$b \rightarrow s \gamma$~\cite{cleo}. The process $b \rightarrow s
\ell \overline{\ell}$ which can only happen at loop level provides a
sensitive and stringent test of the unitarity
of the 
CKM mixing matrix. Thus, it can serve as a test for the validity of
the SM.

The decay process $B_c \rightarrow D_s(1968) \ell \overline{\ell}$
has been studied in a number of models, such as quantum
chromodynamics (QCD) sum rules~(SR) and relativistic quark
model~(RQM) \cite{Azizi,geng,Choi:2010ha,para1}. However, there are
few investigations on the decay $B_c\rightarrow
D_s^*(2317)\ell\overline{\ell}$ \cite{Ghahramany}. The $D_s^*(2317)$
meson is considered to be of controversial since it was discovered
in BaBar \cite{babar}. It was predicted to be broad and available to
decay into $DK$ and $D^*K$ in the potential-based quark
models~\cite{godfreya,godfreyb}. However, the BaBar results show
that $D_s^*(2317)$ is below the $DK$ and $D^*K$ thresholds and has a
narrow decay width. Many works were done to clarify this disparity
between theories and experiments. Some physicists advocated that
$D_s^*(2317)$ is a $DK$ molecular ~\cite{barnes}, a $D_s \pi$
atom~\cite{Szczepaniak} or a four-quark bound
state~\cite{fourquark}, but some studies based on the heavy quark
effective theory (HQET)~\cite{bardeena,bardeenb} suggested that it
is a conventional $c\bar{s}$ state. Following
Ref.~\cite{bardeena,bardeenb}, we suppose that $D_s^*(2317)$ is a
$c\bar{s}$ scalar meson with even parity. The study of form factors
for $B_c \rightarrow D_s^*(2317) \ell \overline{\ell}$ process can
also help us to learn more about the structure of $D_s^*(2317)$.

We choose the light-cone quark model~(LCQM)~\cite{brodsky,BHL,BPP}
to perform the calculation in our work. LCQM takes an advantage of
the equal light-cone time~($\tau = t + z/c$) quantization and
includes the important relativistic effects which are neglected in
the traditional constituent quark model. In addition, compared with
the complex vacuum in equal-time QCD, the vacuum in light-cone
coordinates is simple, because the Fock vacuum state is the exact
eigenstate of the full hamiltonian and all constituents in a
physical eigenstate are directly related to that state. LCQM was
widely used in the investigation of hadronic
decays~\cite{CHY,HYcheng,Choi,yujh} and electromagnetic transition
form factors~\cite{huangtaoa,huangtaob}, and it was proved
successful in explaining the experimental data. We calculate the
form factors, branching ratios and longitudinal lepton polarization
asymmetries~(LPAs) for the two decay processes in the framework of
LCQM and compare the results with each other.

This paper is organized as follows. In Sect.~2, we discuss the
standard model effective hamiltonian for $b\rightarrow s \ell
\overline{\ell}$ decay. In Sect.~3, we calculate the hadronic form
factors for the two decay processes in the light-cone framework. In
Sect.~4 we present our numerical results. In Sect.~5, we give the
discussion and conclusion.

\section{Effective hamiltonian and form factors}

The rare decay $B_c \rightarrow D_s \ell \overline{\ell}$ is
described by $b \rightarrow s \ell \overline{\ell}$ transition at
quark level. After integrating out heavy top quark and 
$W^\pm$ bosons, one can write the effective interacting hamiltonian
density responsible for this transition as~\cite{gws}:
\begin{eqnarray}\label{hamiltonian}
\mathcal{H}(b\rightarrow s \ell^+{\ell}^-)&=&\frac{G_F\alpha}{\sqrt{2}\pi}V_{tb}V_{ts}^*\left[C_9^{\mathrm{eff}}(m_b)\bar{s}_L\gamma_{\mu}b_L\overline{\ell}\gamma^{\mu}\ell\right.\nonumber\\
&&\left.-\frac{2m_bC_7(m_b)}{q^2}\bar{s}_Li\sigma_{\mu\nu}q^{\nu}b_R\overline{\ell}\gamma^{\mu}\ell\right.\nonumber\\&+&\left.C_{10}(m_b)\bar{s}_L\gamma_{\mu}b_L\overline{\ell}\gamma^{\mu}\gamma_5\ell\right],\nonumber\\
\mathcal{H}(b\rightarrow s
\nu\bar{\nu})&=&\frac{G_F}{\sqrt{2}}\frac{2\alpha
V_{tb}V_{ts}^*}{\pi
\textrm{sin}^2\theta_W}X(x_t)\bar{b}\gamma_{\mu}P_Ls\bar{\nu}_l\gamma^{\mu}P_L\nu_l,\nonumber\\
\end{eqnarray}
where $G_F$ is the Fermi constant, $\alpha$ is the electromagnetic
fine structure constant and $V_{ij}$ are the CKM matrix elements.
$X(x_t)$, the top quark loop function, is given by:
\begin{eqnarray}
X(x_t)=\frac{x_t}{8} \frac{(2 + x_t)}{(x_t - 1)} + \frac{(3x_t -
6)}{(1 - x_t)^2}\textrm{ln}x_t, \quad
\big(x_t=\frac{M_t^2}{M_W^2}\big)\nonumber
\end{eqnarray}
and $C_i(\tilde{{\mu}})$ are the Wilson coefficients. In particular,
$C_9^{\mathrm{eff}}$, defined as an effective coefficient and
containing the contribution from the charm-loop, is given
by~\cite{c9}:
\begin{eqnarray}
C_9^{\mathrm{eff}}(\tilde{\mu})&=&C_9+(3C_1+C_2+ 3C_3+ C_4 + 3C_5
\nonumber\\&+&
C_6)h(\hat{m}_c,\hat{s})-\frac{1}{2}h(0,\hat{s})(C_3+3C_4)\nonumber\\
&-&\frac{1}{2}h(1,\hat{s})(4C_3+4C_4
+3C_5+C_6)\nonumber\\&+&\frac{2}{9}(3C_3 + C_4 + 3C_5 + C_6),
\end{eqnarray}
where
\begin{eqnarray}\label{Eq:h}
h(\hat{m}_q,\hat{s})&=&-\frac{8}{9}\textmd{ln}\frac{m_b}{\tilde{\mu}}-\frac{8}{9}\textmd{ln}\hat{m}_q+\frac{8}{27}+\frac{4}{9}x-\frac{2}{9}(2+x)\nonumber\\&\times&|1-x|^{1/2}
\big[\Theta(1-x)\big(\textmd{ln}|\frac{\sqrt{1-x}+1}{\sqrt{1-x}-1}|-i\pi\big)\nonumber\\
&+&\Theta(x-1)2\arctan\frac{1}{\sqrt{x-1}}\big],
\end{eqnarray}
\begin{eqnarray}
 h(0,\hat{s})&=&-\frac{8}{9}\textmd{ln}\frac{m_b}{\tilde{\mu}}+\frac{8}{27}-\frac{4}{9}\textmd{ln}\hat{s}+\frac{4}{9}i\pi,
\end{eqnarray}
in which $\hat{s}=q^2/m_b^2$, $\hat{m}_q=m_q/m_b$ and
$x=4m_q^2/q^2$, where $m_q$ is the constituent quark mass.

In Eq.~(\ref{Eq:h}), we neglect long distance contributions from
charmonia vector resonances $J/\Psi, \Psi'$,...~\cite{Choi,LDa,LDb}.
To evaluate the decay rates and other physical quantities with this
effective hamiltonian, we write the matrix elements of the effective
nontrivial vertexes in Eq.~(\ref{hamiltonian}) in terms of hadronic
form factors:
\begin{eqnarray}
\langle D_s(1968)|\bar{s}\gamma_{\mu} b|B_c\rangle
&=&[f_+(q^2)P_{\mu}+f_-(q^2)q_{\mu}],\\ \langle
D_s(1968)|i\bar{s}\sigma_{\nu\mu}\gamma_5
bq^{\nu}|B_c\rangle&=&\frac{1}{M_{B_c}+M_{D_s}} \\
&&\times\left[q^2P_{\mu}-(P\cdot q)q_{\mu}\right]F_T(q^2), \nonumber
\end{eqnarray}
\begin{eqnarray}
\langle D_s^*(2317)|i\bar{s}\gamma_{\mu}\gamma_5
b|B_c\rangle&=&-[u_+(q^2)P_{\mu}+u_-(q^2)q_{\mu}], ~~~~\\
\langle D_s^*(2317)|\bar{s}\sigma_{\nu\mu}\gamma_5
bq^{\nu}|B_c\rangle&=&\frac{1}{M_{B_c}+M_{D_s^*}}\\
&&\times[q^2P_{\mu}-(P\cdot q)q_{\mu}]U_T(q^2), \nonumber
\end{eqnarray}
where $P=P_{B_c}+P_{D_s}$ and $q=P_{B_c}-P_{D_s}$. These form
factors defined above are related to the commonly used
Bauer-Stech-Wirbel (BSW) form factors~\cite{BSW} via:
\begin{eqnarray}
F_1^{PP}(q^2)&=&f_+(q^2),\nonumber\\
F_0^{PP}(q^2)&=&f_+(q^2)+\frac{q^2}{M_{B_c}^2-M_{D_s}^2}f_-(q^2),\\
F_1^{PS}(q^2)&=&u_+(q^2),\nonumber\\
F_0^{PS}(q^2)&=&u_+(q^2)+\frac{q^2}{M_{B_c}^2-M_{D^*_s}^2}u_-(q^2).
\end{eqnarray}

Then the differential decay rate of the exclusive processes
$B_c\rightarrow D_s(1968)\ell\overline{\ell}$ can be expressed in
terms of the form factors as:
\begin{eqnarray}\label{deff rate}
\frac{\mathrm{d\Gamma}(B_c\rightarrow D_s \ell^+\ell^-)}{\mathrm{d}\hat{s}}&=&\frac{G_F^2M_B^5\alpha^2}{3\cdot2^9\pi^5}|V_{ts}^*V_{tb}|^2\hat{\phi}^\frac{1}{2}(1-4\frac{m_l^2}{q^2})^{\frac{1}{2}}\nonumber\\
&\times&\left[\hat{\phi}\left(1+\frac{2m_l^2}{q^2}\right)f_{T+}+6\frac{m_l^2}{q^2}f_{0+}\right],\nonumber\\
\frac{\mathrm{d\Gamma}(B_c\rightarrow D_s \nu\bar{\nu})}{\mathrm{d}\hat{s}}&=&\frac{G_F^2M_B^5\alpha^2}{2^8\pi^5 \textrm{sin}\theta_W^4}|X(x_t)|^2|V_{ts}^*V_{tb}|^2\hat{\phi}^{\frac{3}{2}}|f_+|^2,\nonumber\\
\end{eqnarray}
where
\begin{eqnarray}
f_{T+}&=&\left|C^{\mathrm{eff}}_9f_+-\frac{2C_7F_T}{1+\sqrt{\hat{r}}}\right|^2+|C_{10}f_+|^2,\nonumber\\
f_{0+}&=&|C_{10}|^2[(1-\hat{r})^2|f_0|^2-\hat{\phi}|f_+|^2],\nonumber\\
\hat{\phi}&=&(\hat{s}-\hat{r}-1)^2-4\hat{r},\nonumber\\
f_0&=&f_++\frac{q^2}{M_{B_c}^2-M_{D_s}^2}f_-,\nonumber\\
\hat{s}&=&q^2/M_{B_c}^2,\quad \hat{r}=M_{D_s}^2/M_{B_c}^2.
\end{eqnarray}

The longitudinal LPAs can be defined as:
\begin{eqnarray}\label{pla}
P_L(\hat{s})=\frac{\mathrm{d\Gamma}_{h=-1}/\mathrm{d}\hat{s}-\mathrm{d\Gamma}_{h=1}/\mathrm{d}\hat{s}}{\mathrm{d\Gamma}_{h=-1}/\mathrm{d}\hat{s}+\mathrm{d\Gamma}_{h=1}/\mathrm{d}\hat{s}},
\end{eqnarray}
where the subscript $h$ is the helicity of the $\ell^{-}$ in final
states. From Eq.(\ref{pla}), we can obtain
that~\cite{GengandKao1996}:
\begin{eqnarray}
P_L(\hat{s})=\frac{2(1-4\frac{m_l^2}{q^2})^{1/2}\hat{\phi}
C_{10}f_+\left[f_+ \textmd{Re}
C_9^{\mathrm{eff}}-\frac{2C_7F_T}{1+\sqrt{\hat{r}}}\right]}{\left[\hat{\phi}\left(1+\frac{2m_l^2}{q^2}\right)F_{T+}+6\frac{m_l^2}{q^2}f_{0+}\right]}.
~~~~\label{plaa}
\end{eqnarray}

For the case of $B_c\rightarrow D_s^*(2317)\ell\overline{\ell}$
processes, we just need to replace the form factors $f_+$, $f_-$ and
$F_T$ in Eq.~(\ref{deff rate}) and Eq.~(\ref{plaa}) with $u_+$,
$u_-$ and $U_T$ respectively.

\section{Form factors in light-cone framework} In LCQM, a meson
can be considered as a quark-antiquark composed system. Assuming a
meson with light-cone momentum
$(P^+,(M^2+P_{\perp}^2)/P^+,\textbf{P}_{\perp})$ is composed of two
constituents $q_1$ and $q_2$, we can give the light-cone components
of the momenta $p_1$ and $p_2$ as:
\begin{eqnarray}
p_1^+=xP^+, & & p_2^+=(1-x)P^+,\nonumber\\
\textbf{p}_{1\perp}=x\textbf{P}_{\perp}+\textbf{k}_{\perp}, & &
\textbf{p}_{2\perp}=(1-x)\textbf{P}_{\perp}-\textbf{k}_{\perp}.
\end{eqnarray}
The light-cone wave function in the momentum space for a
$^{2S+1}L_J$ meson is given by:
\begin{eqnarray}\label{Eq:phi}
\Psi^{JJ_z}_{LS}=\frac{1}{\sqrt{N_c}}\langle
LS;L_z,S_z|LS;J,J_z\rangle \nonumber\\ \times
R^{SS_Z}_{\lambda_1\lambda_2}(x,\textbf{p}_\perp)\varphi_{LL_Z}(x,\textbf{p}_\perp),
\end{eqnarray}
where $\langle LS;L_z,S_z|LS;J,J_z\rangle$ are the Clebsch-Gordon
coefficients and $R^{SS_Z}_{\lambda_1\lambda_2}(x,\textbf{p}_\perp)$
are the Melosh
transformation~\cite{bqma1,bqma2,bqma3,bqma4,melosh1,melosh2} matrix
elements, which account for the relativistic effect due to quark
transversal motions inside hadrons. Such an effect plays an
important role to understand the famous proton ``spin
puzzle"~\cite{MaOld1,MaOld2}.

We use Gaussian-type wave functions~\cite{BHL} to describe the
radial part $\varphi_{LL_Z}(x,\textbf{p}_\perp)$ :
\begin{eqnarray}\label{Eq:wf1}
\varphi(x,\textbf{p}_{\perp})_{L=0}&=&\frac{4\pi^{3/4}}{\beta^{3/2}}\sqrt{\frac{\mathrm{d}p_z}{\mathrm{d}x}}\textrm{exp}(-\frac{\textbf{p}_{\perp}^2+p_z^2}{2\beta^2}),
\end{eqnarray}
\begin{eqnarray}\label{Eq:wf2}
\varphi(x,\textbf{p}_{\perp})_{L=1}&=&\frac{4\sqrt{2}\pi^{3/4}}{\beta^{5/2}}\sqrt{\frac{\mathrm{d}p_z}{\mathrm{d}x}}p_{L_z}\textrm{exp}(-\frac{\textbf{p}_{\perp}^2+p_z^2}{2\beta^2}),\nonumber\\
\end{eqnarray}
where
\begin{eqnarray}
p_{L_z=\pm1}&=&\frac{\mp(p_x\pm ip_y)}{\sqrt{2}}, ~~ p_{L_z=0}=p_z.
\end{eqnarray}
In 
the light-cone framework, $p_z$ can be represented as:
\begin{eqnarray}
p_z=(x-\frac{1}{2})M_0+\frac{m_2^2-m_1^2}{2M_0},
\end{eqnarray}
where $M_0^2=\sum^2_{i=1}(\textbf{k}^2_{\perp i}+m_i^2)/x_i$ and
$m_i$ is the constituent quark mass.

For pseudoscalar mesons 
$({}^{2S+1}L_J={}^1S_0)$, the spin-orbit part
$R^{SS_Z}_{\lambda_1\lambda_2}(x,\textbf{p}_\perp)$ can be
simplified as an effective vertex form:
\begin{eqnarray}
R^{00}_{\lambda_1\lambda_2}(x,\textbf{p}_\perp)=-\frac{\overline{u}(p_1,\lambda_1)\gamma_5
v(p_2,\lambda_2)}{\sqrt2 \widetilde{M}_0},\nonumber\\
\end{eqnarray}
where $ \widetilde{M}_0=\sqrt{M_0^2-(m_1-m_2)^2}$.

Correspondingly, for scalar mesons 
$({}^{2S+1}L_J={}^3P_0)$, we can
also write 
an effective vertex by combining Clebsch-Gordon coefficients,
spin-orbit part $R^{SS_Z}_{\lambda_1\lambda_2}(x,\textbf{p}_\perp)$
and $p_{L_z}$ in the radial part
$\varphi(x,\textbf{p}_{\perp})_{L=1}$ together as~\cite{HYcheng}:
\begin{eqnarray}
\langle 1S;L_z,S_z|1S;J,J_z\rangle
R^{1S_z}_{\lambda_1\lambda_2}(x,p_\perp)p_{L_z}\nonumber\\=i\overline{u}(p_1,\lambda_1)
v(p_2,\lambda_2)\frac{\widetilde{M}_0}{\sqrt6 M_0}.
\end{eqnarray}

In LCQM, the Drell-Yan-West~(DYW) ($q^+=0$) frame~\cite{dyw1,dyw2}
is widely used to calculate form factors. We can avoid the
non-valence diagrams arising from the quark-antiquark pair creation
(so-called ¡°Z-graph¡±)~\cite{zgraph} by choosing DYW frame. In this
frame, the momenta of mesons in the initial and final states are
represented as:
\begin{eqnarray}
q=\big(0,\frac{\textbf{q}^2}{P^+},\textbf{q}_{\perp}\big),
P_{B_c}=\big(P^+,\frac{M_{B_c}^2}{P^+},\textbf{0}\big),\nonumber\\
P_D=\big(P^+,\frac{M_D^2+\textbf{q}_{\perp}^2}{P^+},-\textbf{q}_{\perp}\big),
\end{eqnarray}
and the momenta of constituent quarks are represented as:
\begin{eqnarray}
p_{\bar{c}}&=&\left(xP^+,\frac{m_{\bar{c}}^2+\textbf{k}_{\perp}^2}{xP^+},-\textbf{k}_{\perp}\right),\nonumber\\
p_b&=&\left((1-x)P^+,\frac{m_b^2+\textbf{k}_{\perp}^2}{(1-x)P^+},\textbf{k}_{\perp}\right),\nonumber\\
p_s&=&\left((1-x)P^+,\frac{m_s^2+(\textbf{k}_{\perp}-\textbf{q}_{\perp})^2}{(1-x)P^+},\textbf{k}_{\perp}-\textbf{q}_{\perp}\right).\nonumber\\
\end{eqnarray}
With the effective vertex and the wave functions given in
Eq.~(\ref{Eq:phi})$\sim$Eq.~(\ref{Eq:wf2}), we can give the explicit
forms of the form factors $f_+(q^2)$, $f_-(q^2$), $F_T(q^2)$,
$u_+(q^2)$, $u_-(q^2)$ and $U_T(q^2)$ (see in appendix).

Noticing that all the form factors are calculated in the space-like
region with $q^2=q^+q^--\textbf{q}_{\perp}^2\leq 0$, while $B_c$
meson rare decays are defined in the time-like region, we need to
parameterize the form factors as explicit functions of $q^2$ in the
space-like region and then extended them through the analytical
continuation to the time-like region. We choose a three-parameter
form in this paper as:
\begin{eqnarray}\label{parameterize}
F(q^2)=F(0)\textrm{exp}[\textrm{a}(q^2/M_{B_c}^2)+\textrm{b}(q^2/M_{B_c}^2)^2],
\end{eqnarray}
where $F(q^2)$ denotes any one of the form factors used in this
paper.

\section{Numerical results}
In this section, we calculate the form factors, branching ratios and
longitudinal LPAs with input parameters. The Wilson coefficients and
other electro-weak constants used in Eq.~(\ref{hamiltonian}) and
Eq.~(\ref{deff rate}) are given in
Table~\ref{tap:wilson}~\cite{para1}:

\begin{table}

\caption{The electro-weak parameters}\label{tap:wilson}
\begin{tabular}{llll}
\hline\noalign{\smallskip} Parameter&Value&Parameter&Value\\
\noalign{\smallskip}\hline\noalign{\smallskip}
$m_W$&80.41 GeV&$C_1$&-0.248\\
$m_Z$&91.837 GeV&$C_2$&1.107\\
$\sin^2\theta_W$&0.2233&$C_3$&0.011\\
$\alpha^{-1}$&129&$C_4$&-0.026\\
$|V_{tb}^*V_{ts}|$&0.0385&$C_5$&0.007\\
$C_6$&-0.031&$C_7$&-0.313\\
$C_9^{\mathrm{eff}}$&4.344&$C_{10}$&-4.669\\
\noalign{\smallskip}\hline
\end{tabular}
\end{table}
The constituent quark masses used in LCQM calculation are chosen
as~\cite{Amsler:2008zzb}:
\begin{eqnarray}
m_s=0.37 \  \textrm{GeV},  ~~~~m_c=1.4 \ \textrm{GeV}, ~~~~m_b=4.8\
\textrm{GeV}. \nonumber
\end{eqnarray}
There is still another important parameter $\beta$ which describes
the momenta distribution of constituent quarks in Eq.~(\ref{Eq:wf1})
and Eq.~(\ref{Eq:wf2}). It can be fixed by meson decay constants as:
\begin{eqnarray}
f_P=2\sqrt{6}\int\frac{\mathrm{d}x\mathrm{d}^2\textbf{k}_{\perp}}{16\pi^3}\frac{\mathcal{A}}{\mathcal{A}^2+k_{\perp}^2}\varphi_s(x,\textbf{k}_{\perp}),\nonumber\\
f_S=2\sqrt{6}\int\frac{\mathrm{d}x\mathrm{d}^2\textbf{k}_{\perp}}{16\pi^3}\frac{m_1(1-x)-m_2x}{\mathcal{A}^2+k_{\perp}^2}\varphi_p(x,\textbf{k}_{\perp}),\nonumber\\
\end{eqnarray}
where $\mathcal{A}=m_s(1-x)+m_bx$, $f_P$ and $f_S$ are the decay
constants of pseudoscalar and scalar mesons, and $\varphi_s$ and
$\varphi_p$ are $s$-wave and $p$-wave functions.

The decay constants of $B_c$, $D_s(1968)$ and $D_s^*(2317)$ mesons
in this paper are employed as $f_{B_c}=400\pm40\ \textrm{MeV}$
~\cite{LCD}, $f_{D_s}=257.8\pm5.9\ \textrm{MeV}$~\cite{Kronfeld} and
$f_{D_s^*}=71\ \textrm{MeV}$~\cite{C.W.Hwang2004}. Then, we can fix
the $\beta$ parameters as: $\beta_{B_c}=0.89\pm0.075$, $
\beta_{D_s}=0.56\pm0.011$ and $\beta_{D_s^*}=0.3376\nonumber$.

As we have mentioned above, the physical energy region for rare
leptonic decays is time-like. For $B_c\rightarrow
D_s(1968)\ell\overline{\ell}$ decay process, the region is
$4m_l^2\leq q^2 \leq (M_{B_c}-M_{D_s})^2=18.56~ \textrm{GeV}^2$, and
for $B_c\rightarrow D_s^*(2317)\ell\overline{\ell}$ decay process,
the region is $4m_l^2\leq q^2 \leq (M_{B_c}-M_{D^*_s})^2=15.67
~\textrm{GeV}^2$. Because the form factors in both time-like and
space-like regions share the same form, we can choose the energy
area in space-like region ranging from -25~GeV to 0~GeV to perform
the light-cone calculation and then extract the parameters $a$, $b$
and $F(0)$ in Eq.~(\ref{parameterize}) with the errors coming from
the uncertainties of $\beta$ parameters, so we can acquire the $B_c$
decay form factors.

With a light-cone calculation and parameters fitting, we list the
parameters in the form factors $f_+$, $f_-$, $F_T$, $u_+$, $u_-$ and
$U_T$ in Table~\ref{tab:formf1} and Table~\ref{tab:formf2}.
\begin{table} \caption{Form factors for
$B_c\rightarrow D_s(1968)\ell \overline{\ell}$ decay process}
\label{tab:formf1}
\begin{tabular}{lllll}
\hline &F(0)& a& b\\\hline
$f_+$&$0.25^{+0.03}_{-0.02}$&$2.94^{+0.31}_{-0.22}$&$0.70^{+0.01}_{-0.06}$\\
$f_-$&$-0.245^{+0.19}_{-0.10}$&$3.05^{+0.23}_{-0.27}$&$0.74^{+0.06}_{-0.02}$\\
$F_T$&$-0.357^{+0.04}_{-0.03}$&$2.91^{+0.19}_{-0.20}$&$0.68^{+0.07}_{-0.05}$\\\hline
\end{tabular}
\end{table}

\begin{table}
\caption{Form factors for $B_c\rightarrow D_s^*(2317)\ell
\overline{\ell}$ decay process}\label{tab:formf2}
\begin{tabular}{lllll}
\hline &F(0)& a& b\\\hline
$u_+$&$0.110^{+0.02}_{-0.01}$&$4.093^{+0.67}_{-0.22}$&$0.895^{+0.01}_{-0.04}$\\
$u_-$&$-0.144^{+0.02}_{-0.01}$&$4.235^{+0.30}_{-0.41}$&$0.988^{+0.01}_{-0.00}$\\
 $U_T$&$-0.194^{+0.01}_{-0.02}$&$4.068^{+0.45}_{-0.38}$&$0.885^{+0.01}_{-0.00}$\\\hline
\end{tabular}

\end{table}
In Figs.~\ref{f+}-\ref{ft}, we show our results of $f_+$, $f_-$ and
$f_T$ for $B_c\rightarrow D_s(1968) \ell \overline{\ell}$ decay
process and compare them with other
predictions~\cite{Azizi,geng,Choi:2010ha}. As shown in the figures,
the absolute magnitudes of the form factors in our results are
slightly larger than those in Azizi's~\cite{Azizi}~(dotted curve),
those in Geng's~\cite{geng}~(dashed curve) and those in
Choi's~\cite{Choi:2010ha}~(dash-dotted curve) at $q^2=0$ point. We
also compare the form factors for decay process $B_c\rightarrow
D_s(1968) \ell \overline{\ell}$ with those for decay process
$B_c\rightarrow D_s^*(2317) \ell \overline{\ell}$ in
Figs.~\ref{plus}-\ref{trans}. We can see from the figures that the
absolute magnitudes of form factors for decay process
$B_c\rightarrow D_s(1968)\ell \overline{\ell}$ are about twice
larger than those for $B_c\rightarrow D_s^*(2317) \ell
\overline{\ell}$ process at $q^2=0$ point, but as $q^2$ become
large, they tend to be the same.

 \begin{center}
      \includegraphics[width=3.2in]{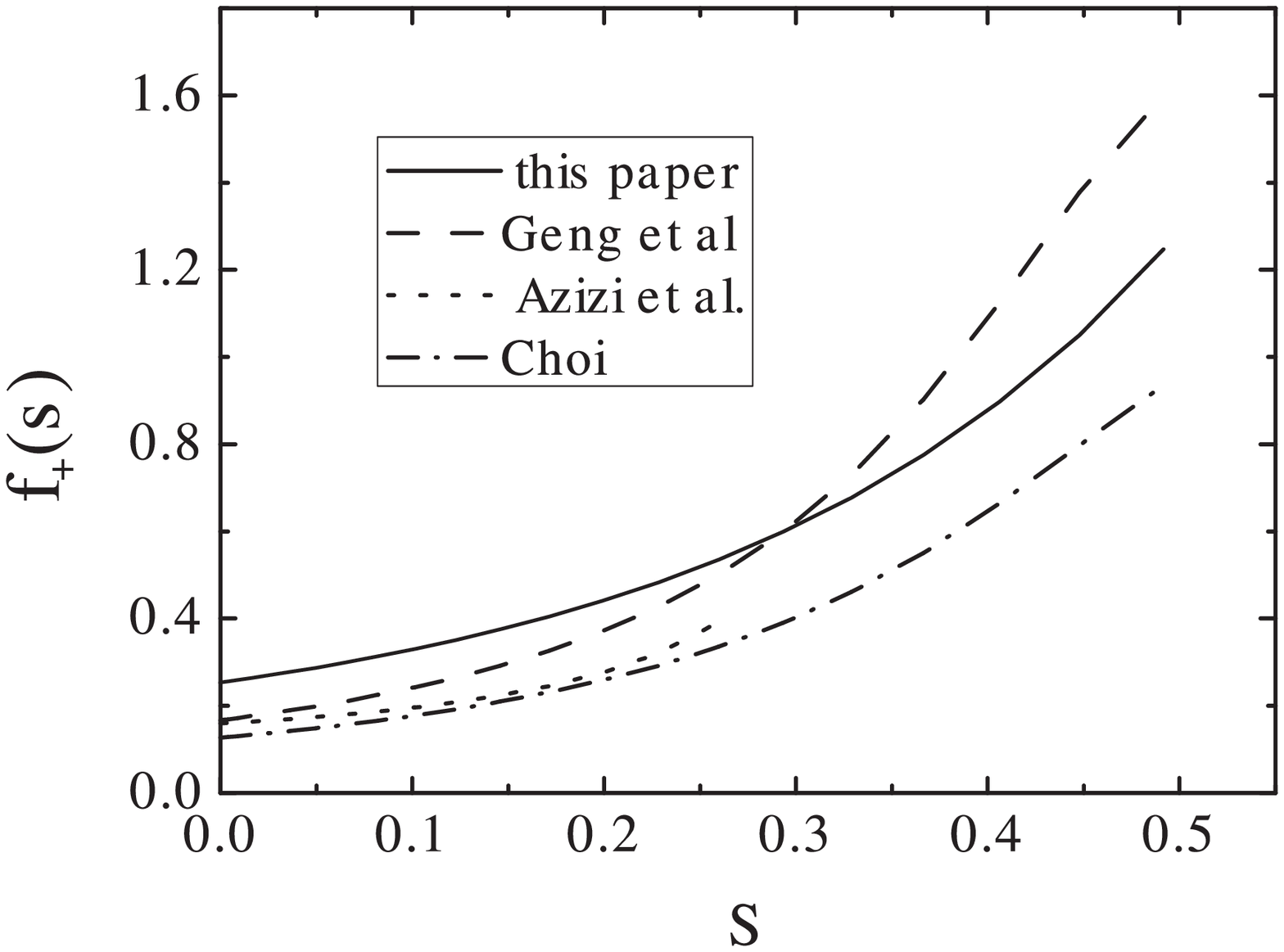}
    \figcaption{$f_+(q^2)$ for $B_c \rightarrow D_s(1968) \ell
    \overline{\ell}$ with definition $S=q^2/M_{B_c^2}$. Our results are represented by \emph{solid curve},
     Azizi's~\cite{Azizi} are represented by \emph{dotted curve}, Geng's~\cite{geng} are represented by \emph{dashed curve} and
Choi's~\cite{Choi:2010ha} are represented by \emph{dash-dotted
curve} respectively.}\label{f+}
\end{center}

 \begin{center}
      \includegraphics[width=3.2in]{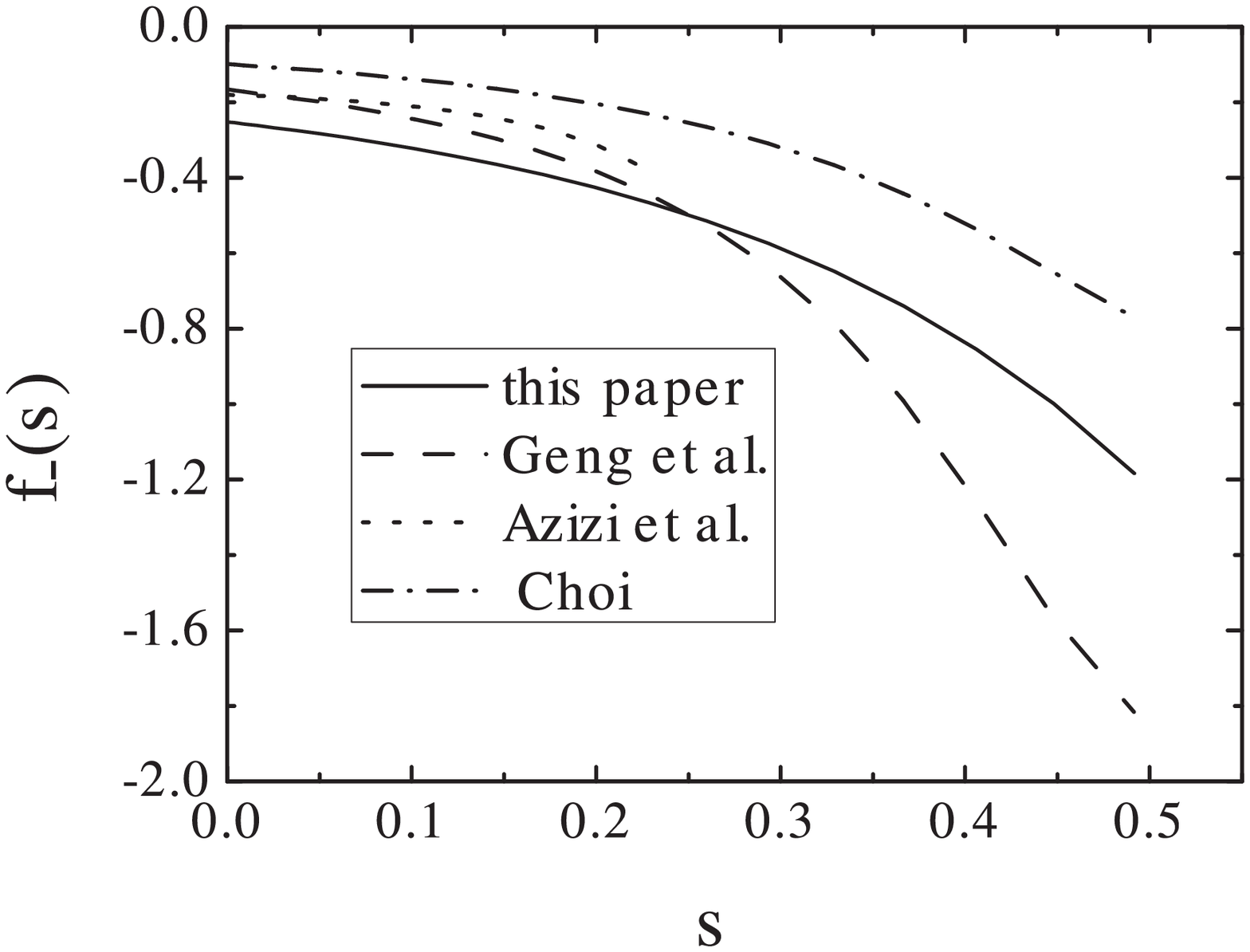}
    \figcaption{$f_-(q^2)$ for $B_c \rightarrow D_s(1968)\ell \overline{\ell}$ process. Our results are represented by the \emph{solid curve},
     Azizi's~\cite{Azizi} are represented by the \emph{dotted curve}, Geng's~\cite{geng} are represented by the \emph{dashed curve} and
Choi's~\cite{Choi:2010ha} are represented by the \emph{dash-dotted
curve} respectively.}\label{f-}
\end{center}

 \begin{center}
      \includegraphics[width=3.2in]{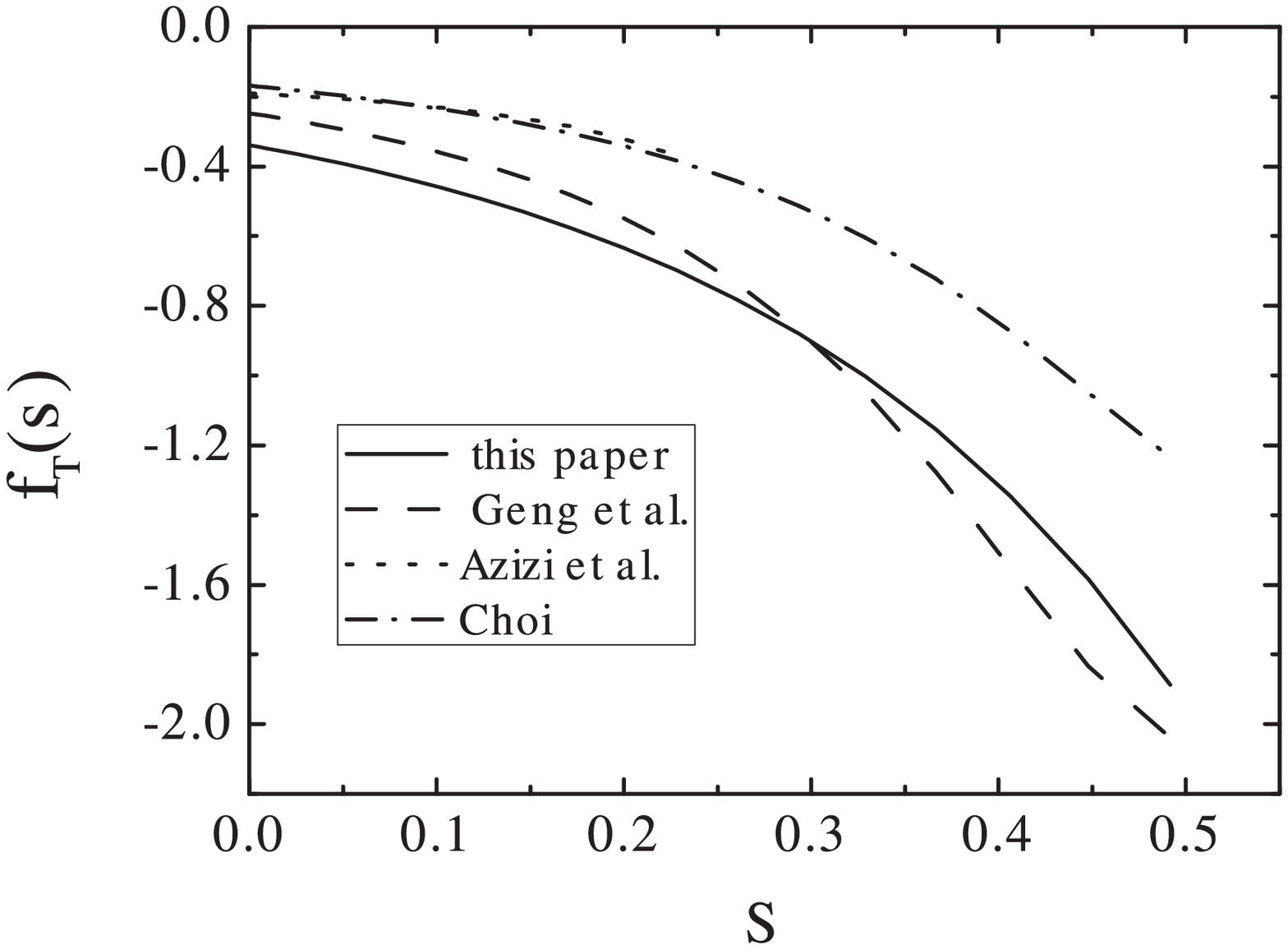}
    \figcaption{$f_T(q^2)$ for $B_c \rightarrow D_s(1968) \ell \overline{\ell}$ process. Our results are represented by the \emph{solid curve},
     Azizi's~\cite{Azizi} are represented by the \emph{dotted curve}, Geng's~\cite{geng} are represented by the \emph{dashed curve} and
Choi's~\cite{Choi:2010ha} are represented by the \emph{dash-dotted
curve} respectively.}\label{ft}
\end{center}

 \begin{center}
      \includegraphics[width=3.2in]{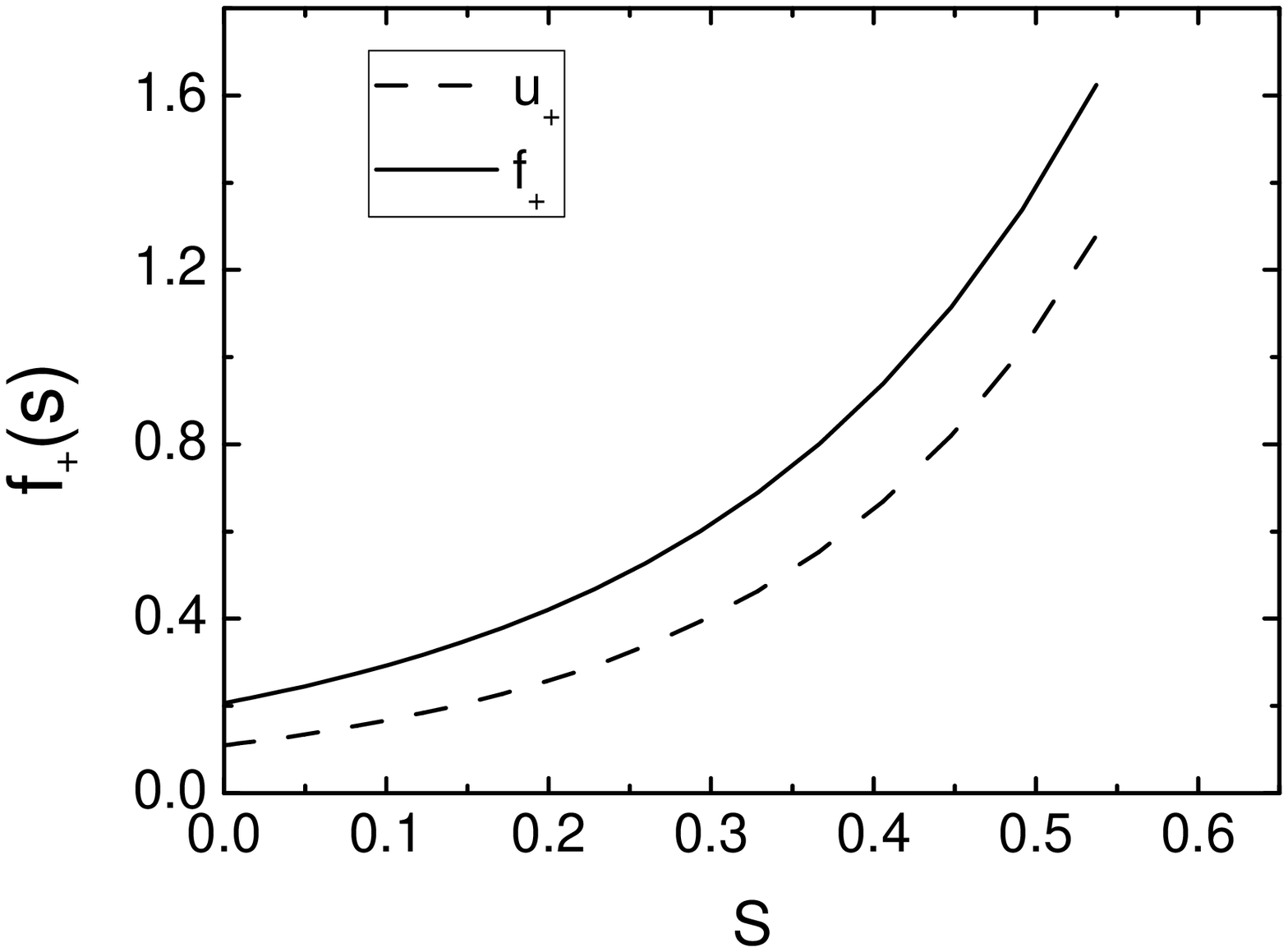}
    \figcaption{$f_+(q^2)$ for $B_c \rightarrow D_s(1968) \ell \overline{\ell}$ process represented by the \emph{solid curve} compared with $u_+(q^2)$
    for $B_c \rightarrow D_s^*(2317) \ell \overline{\ell}$ process represented by the \emph{dashed curve}.}\label{plus}
\end{center}

 \begin{center}
      \includegraphics[width=3.2in]{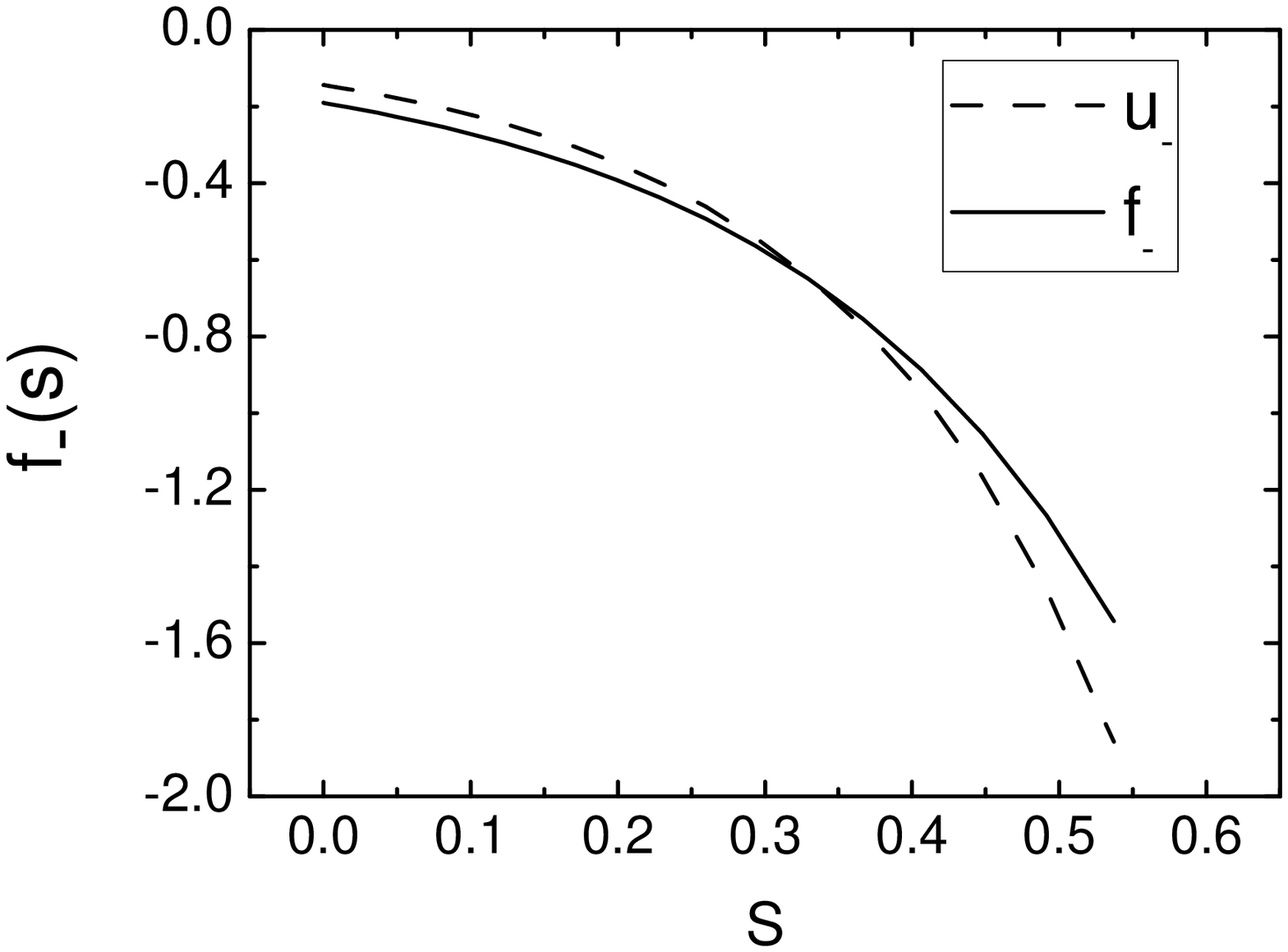}
    \figcaption{$f_-(q^2)$ for $B_c \rightarrow D_s(1968) \ell \overline{\ell}$ process represented by the \emph{solid curve} compared with $u_-(q^2)$
    for $B_c \rightarrow D_s^*(2317) \ell \overline{\ell}$ process represented by the \emph{dashed curve}.}\label{minus}
\end{center}

 \begin{center}
      \includegraphics[width=3.2in]{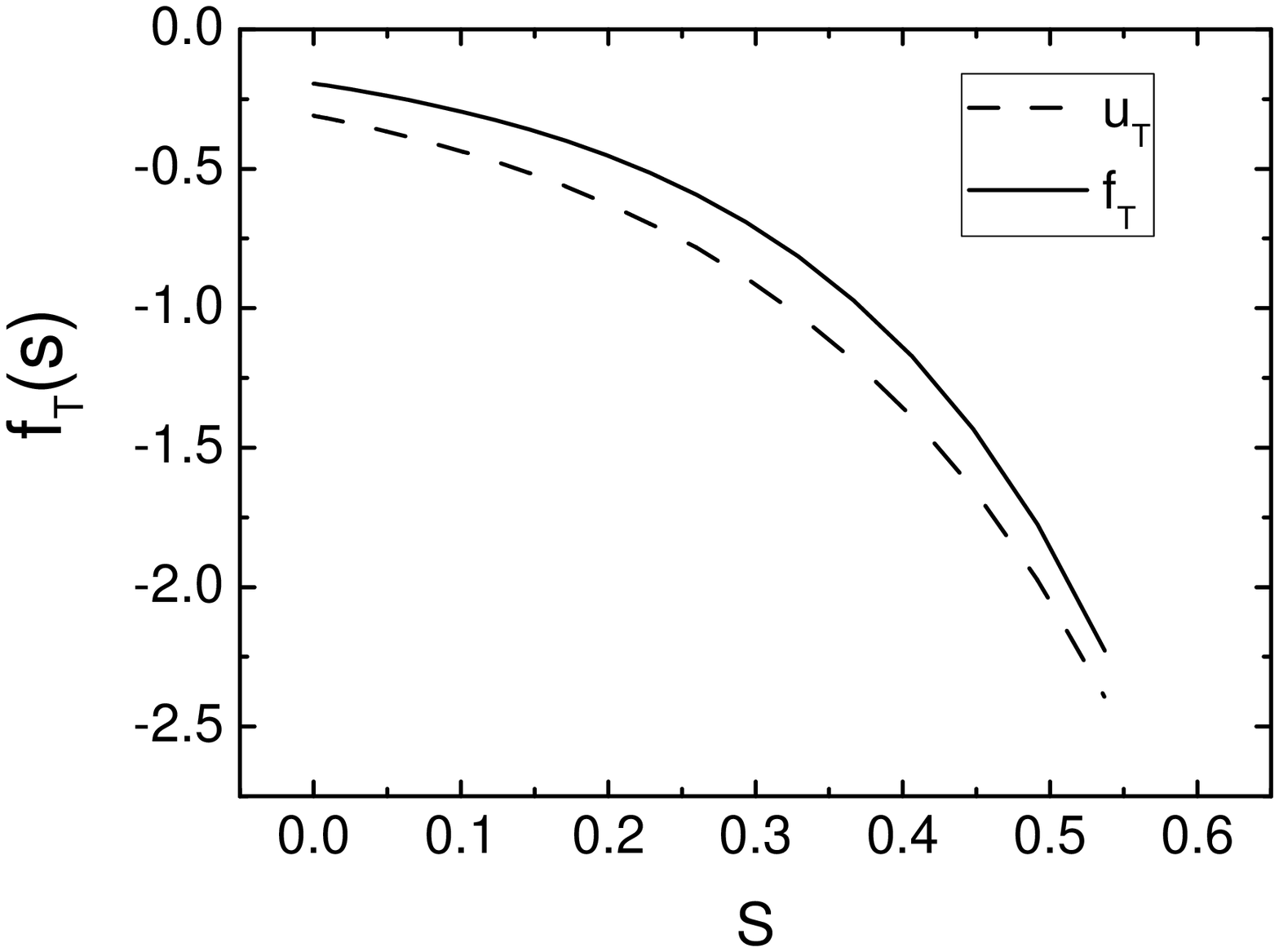}
    \figcaption{$f_T(q^2)$ for $B_c \rightarrow D_s(1968) \ell \overline{\ell}$ process represented by the \emph{solid curve} compared with $u_T(q^2)$
    for $B_c\rightarrow D_s^*(2317) \ell \overline{\ell}$ process represented by the \emph{dashed curve}.}\label{trans}
\end{center}

\begin{table*}
\caption{Branching ratios without long distance contributions for
$B_c\rightarrow D_s(1968)\ell \overline{\ell}$ decay
}\label{tab:br1}
\begin{tabular}{p{4cm}p{3cm}p{3cm}p{3cm}p{3cm}}
\hline &our results&
Azizi~\cite{Azizi}&Geng~\cite{geng}&Choi~\cite{Choi:2010ha}\\\hline
$\mathcal{B}(B_c\rightarrow D_s
\nu\bar{\nu})$&$1.67^{+0.70}_{-0.39}\times 10^{-6}$&$0.49\times
10^{-6}$&$0.92\times 10^{-6}$&$0.37\times
10^{-6}$\\
$\mathcal{B}(B_c\rightarrow D_s
\mu^+\mu^-)$&$1.26^{+1.07}_{-0.30}\times 10^{-7}
$&$0.61\times 10^{-7}$&$1.36\times 10^{-7}$&$0.51\times 10^{-7}$\\
$\mathcal{B}(B_c\rightarrow D_s
\tau^+\tau^-)$&$0.37^{+0.55}_{-0.24}\times
10^{-7}$&$0.23\times 10^{-7}$&$0.37\times 10^{-7}$&$0.13\times 10^{-7}$\\
\hline
\end{tabular}

\end{table*}

\begin{table*}[htbp]

\caption{Branching ratios without long distance contributions for
$B_c\rightarrow D_s^*(2317)\ell \overline{\ell}$ decay
}\label{tab:br2}
\begin{tabular}{p{3cm}p{2.5cm}p{2.8cm}}
\hline &our results&SR~\cite{Ghahramany}\\\hline
$\mathcal{B}(B_c\rightarrow D_s^*
\nu\bar{\nu})$&$3.08^{+2.46}_{-0.73}\times 10^{-7}$
&$(3.06\pm0.76)\times 10^{-7}$\\
$\mathcal{B}(B_c\rightarrow D_s^*
\mu^+\mu^-)$&$2.27^{+1.92}_{-0.57}\times
10^{-8}$&$(3.76\pm 0.92)\times 10^{-8}$ \\
$\mathcal{B}(B_c\rightarrow D_s^*
\tau^+\tau^-)$&$3.53^{+7.30}_{-2.16}\times 10^{-9}$&$(1.28\pm
0.32)\times 10^{-9}$\\ \hline
\end{tabular}
\end{table*}
Differential branching ratios for decay processes $B_c\rightarrow
D_s^*(2317) \ell \overline{\ell}$ and $B_c\rightarrow D_s(1968) \ell
\overline{\ell}$ are shown in Figs.~\ref{nu}-\ref{tau}. We only take
into account the short distance effect in the effective hamiltonian,
so there are no peaks at $c\bar{c}$ resonance threshold. It is
interesting to notice that the form factors for the two decay
processes show few differences, but the differential branching
ratios of the two decay modes have large discrepancies as shown in
Figs.~\ref{nu}-\ref{tau}. The maximum values of differential
branching ratios for $B_c\rightarrow D_s(1968) \ell \overline{\ell}$
decay process are about 3$\sim$10 times larger than those for
$B_c\rightarrow D_s^*(2317) \ell \overline{\ell}$ decay process.
Longitudinal LPAs are shown in Fig.~\ref{plmu} and Fig.~\ref{pltau}.
It is easy to find from Fig.~\ref{plmu} that the LPAs for both
$B_c\rightarrow D_s(1968) \mu^+ \mu^-$ and $B_c\rightarrow
D_s^*(2317) \mu^+ \mu^-$ decay processes are close to -1 in most of
the energy region, and become zero sharply at the end points of $S$.
It can be explained by a formula~\cite{Burdman}:
\begin{eqnarray}
P_L\simeq
\frac{2C_{10}\textrm{Re}C^{\mathrm{eff}}_9}{|C^{\mathrm{eff}}_9|^2+|C_{10}^2|}\simeq-1,
\end{eqnarray}
when lepton mass $m_l\rightarrow 0$.
However, for the case of $\ell=\tau$, because of the heavy mass of
$\tau$, the values of LPAs change remarkably with the variation of
$S$ as shown in Fig.~(\ref{pltau}).

\begin{center}
      \includegraphics[width=3.5in]{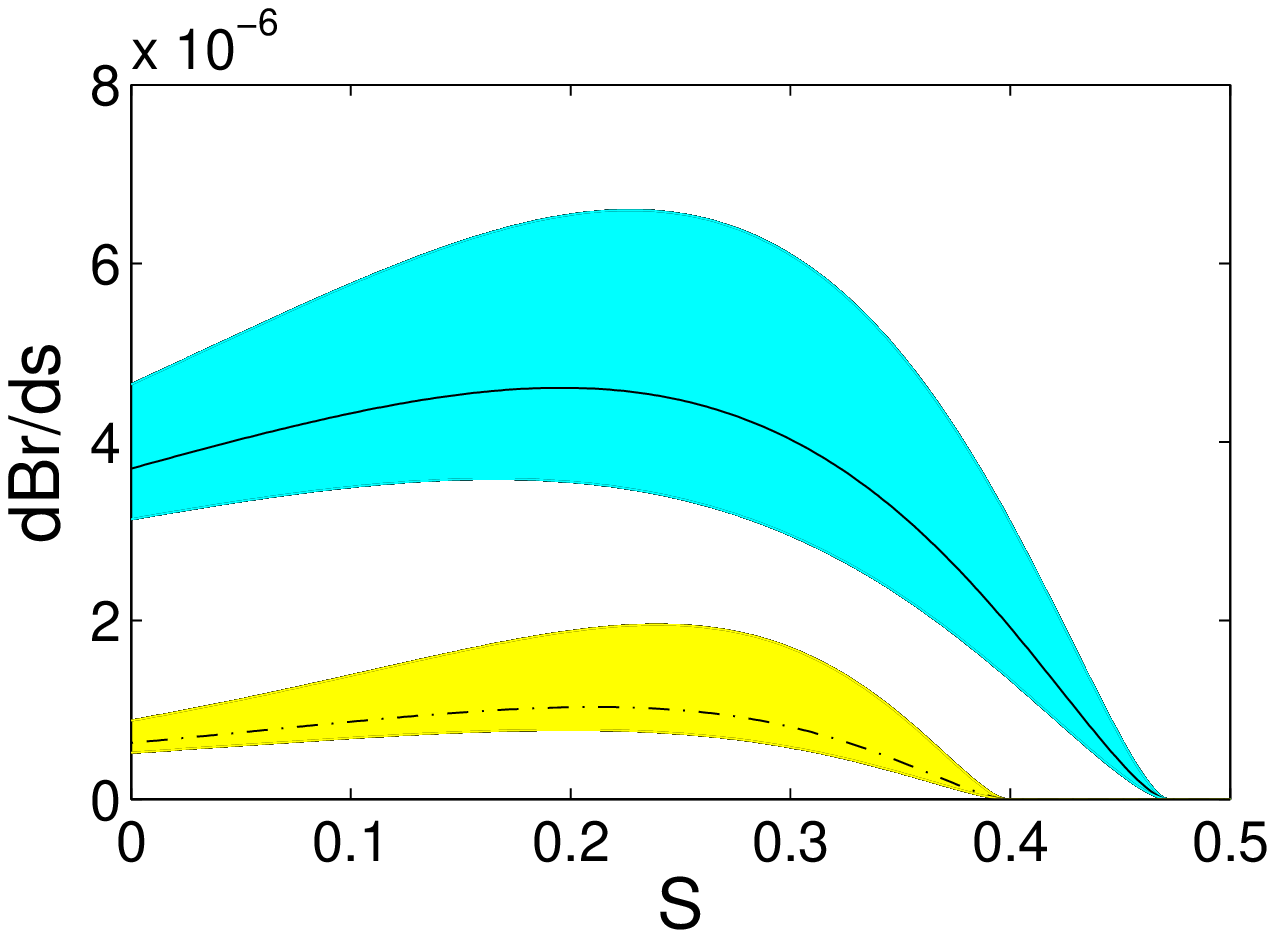}
    \figcaption{Differential branching ratios of $B_c\rightarrow D_s\nu\overline{\nu}$ decay process, represented by the \emph{solid curve},
    and $B_c\rightarrow D_s^*\nu\overline{\nu}$ decay process, represented by the \emph{dash-dotted curve}. The shaded regions show the errors.}\label{nu}
      \includegraphics[width=3.5in]{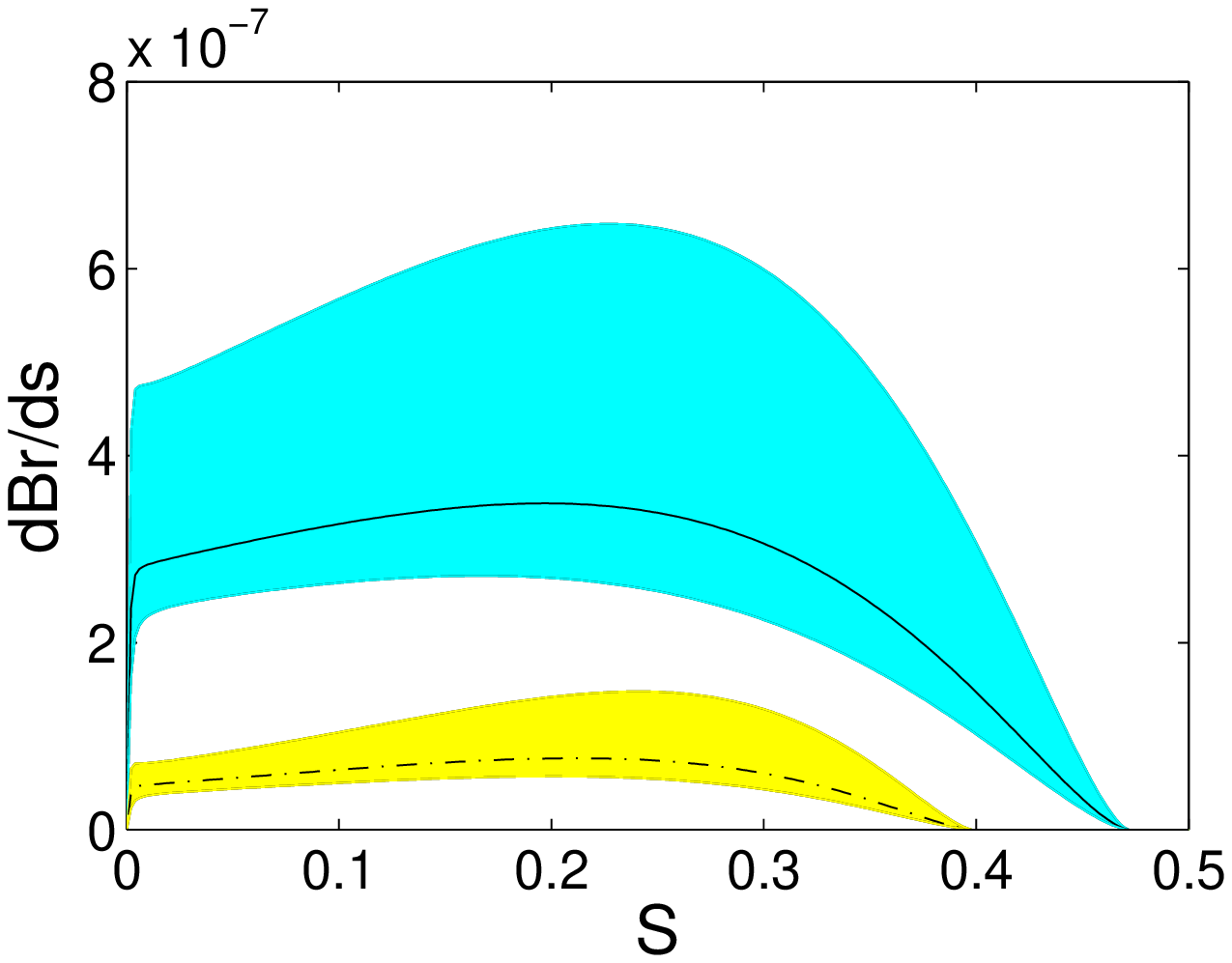}
    \figcaption{Differential branching ratios of $B_c\rightarrow D_s\mu^+\mu^-$ decay process, represented by the \emph{solid curve},
    and $B_c\rightarrow D_s^*\mu^+\mu^-$ decay process, represented by the \emph{dash-dotted curve}. The shaded regions show the errors.}\label{mu}
\end{center}

\begin{center}
      \includegraphics[width=3.4in]{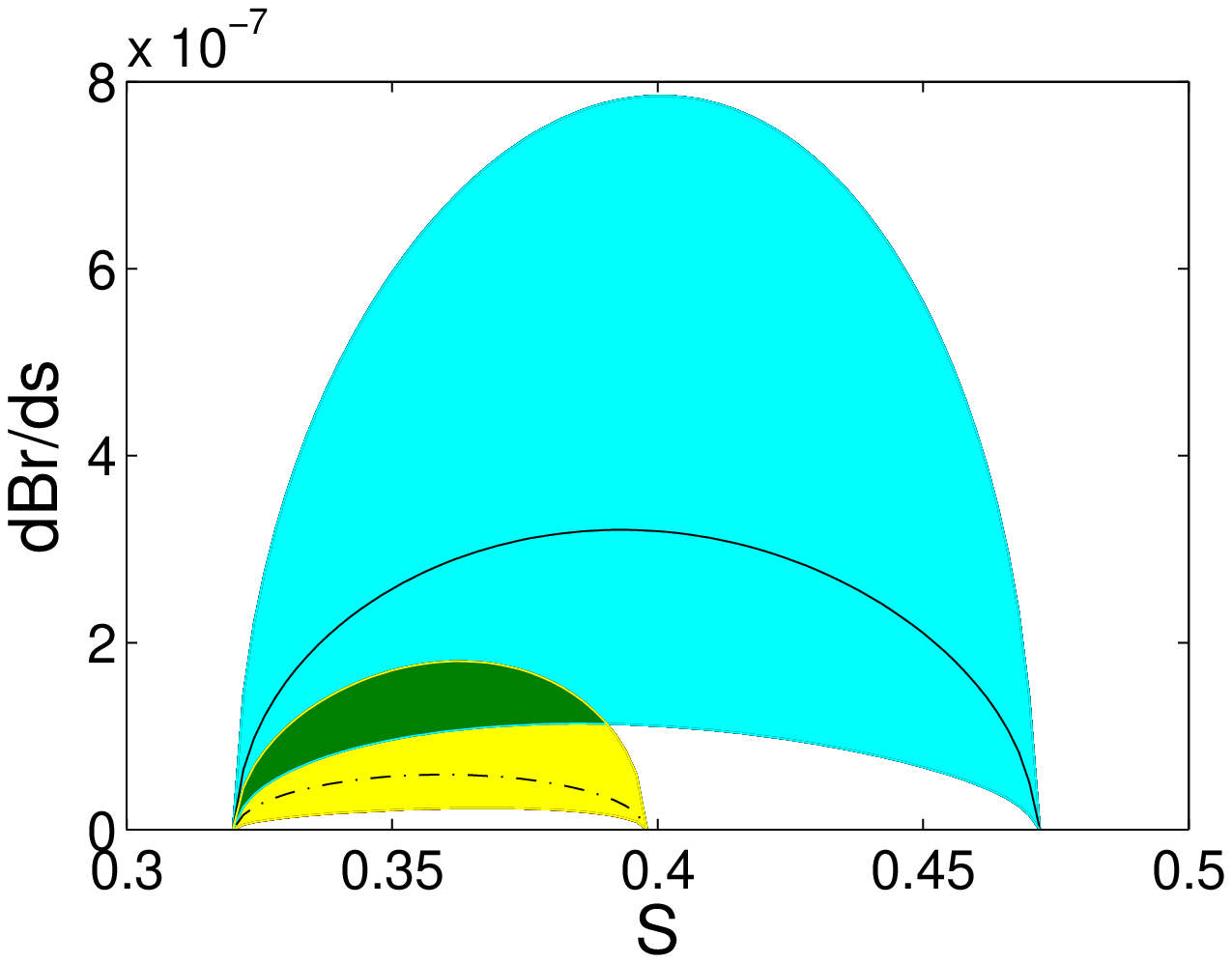}
    \figcaption{Differential branching ratios of $B_c\rightarrow D_s\tau^+\tau^-$ decay process, represented by the \emph{solid curve},
    and $B_c\rightarrow D_s^*\tau^+\tau^-$ decay process, represented by the \emph{dash-dotted curve}. The shaded regions show the errors.}\label{tau}
          \includegraphics[width=3.4in]{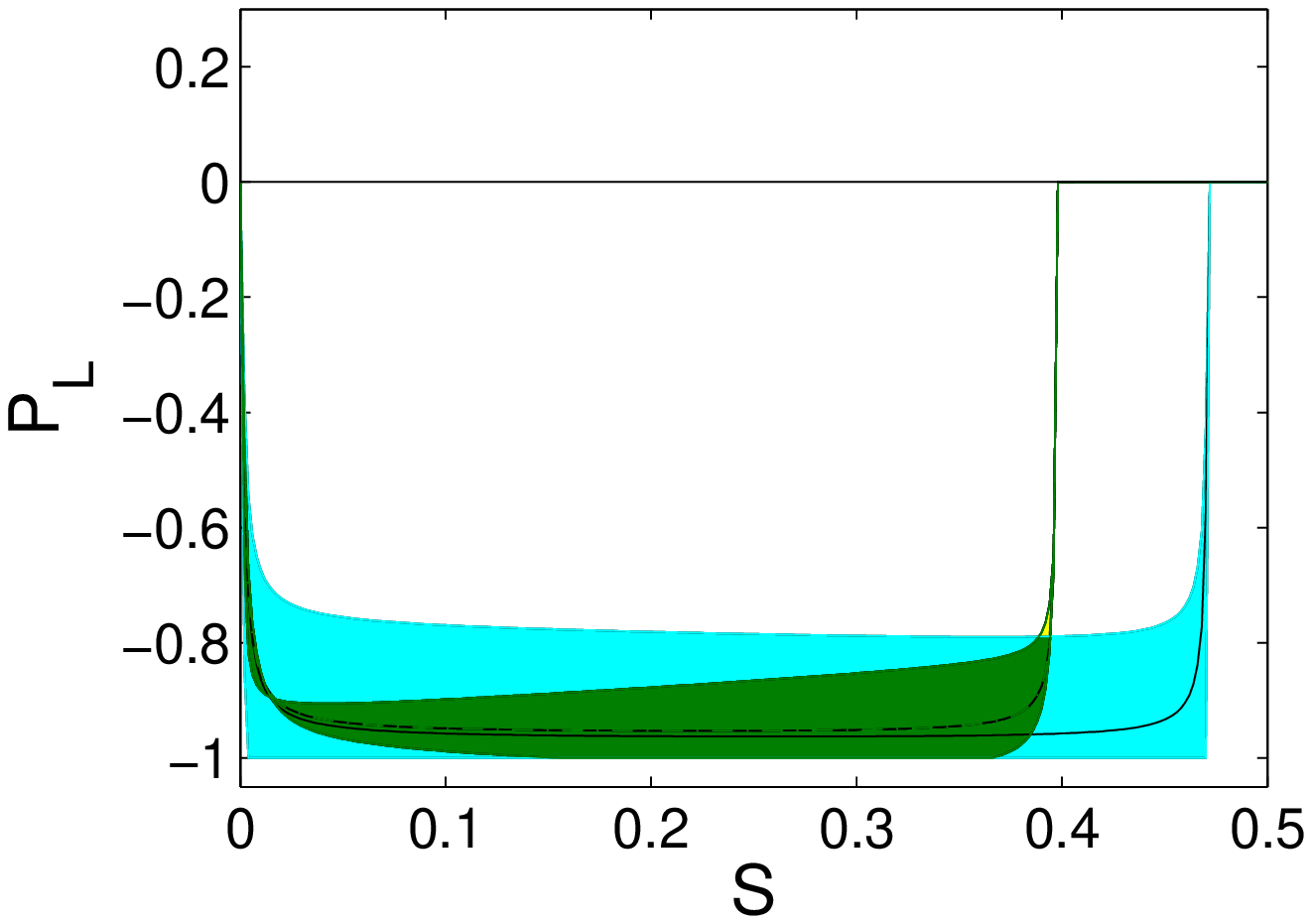}
    \figcaption{Longitudinal lepton polarization
asymmetries of $B_c\rightarrow D_s\mu^+\mu^-$ decay process,
represented by the \emph{solid curve}, and $B_c\rightarrow
D_s^*\mu^+\mu^-$ decay process, represented by the \emph{dash-dotted
curve}. The shaded regions show the errors.}\label{plmu}
\end{center}

\begin{center}
      \includegraphics[width=3.3in]{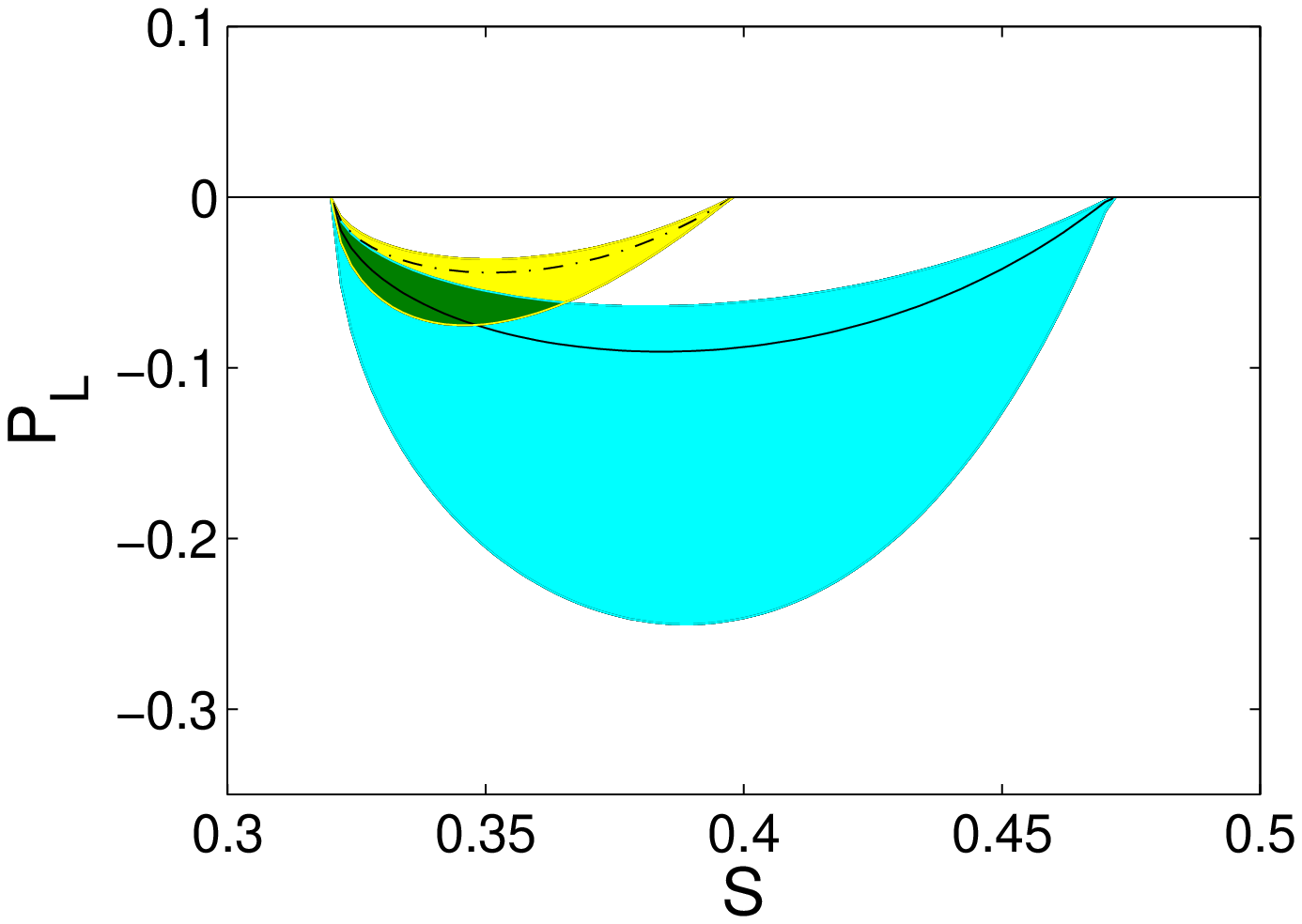}
    \figcaption{Longitudinal lepton polarization
asymmetries of $B_c\rightarrow D_s\tau^+\tau^-$ decay process,
represented by the \emph{solid curve}, and $B_c\rightarrow
D_s^*\tau^+\tau^-$ decay process, represented by the
\emph{dash-dotted curve}. The shaded regions show the
errors.}\label{pltau}
\end{center}

 By integrating the
differential ratios over $S=q^2/M_{B_c}^2$, we can obtain the
branching ratios for the two decay processes. We list the results in
Table~\ref{tab:br1} and Table~\ref{tab:br2} and compare our results
with other predictions.

The average values of LPAs can also be acquired by integral. 
 For the decay process $B_c\rightarrow D_s(1968) \ell^+ \ell^-$,
the average values of $P_L$ are $-0.94^{+0.17}_{-0.06}$,
$-0.07^{+0.01}_{-0.11}$ for $\ell=\mu$,$\tau$ respectively. For the
decay process $B_c\rightarrow D_s^*(2317) \ell^+ \ell^-$, the
average values of $P_L$ are $-0.93^{+0.07}_{-0.04}$,
$-0.032^{+0.006}_{-0.020}$ for $\ell=\mu$,$\tau$ respectively.

\section{Discussion and conclusion}
In this work, we analyzed the rare leptonic decay processes
$B_c\rightarrow D_s(1968)\ell \overline{\ell}$ and $B_c\rightarrow
D_s^*(2317)\ell \overline{\ell}$ within the framework of the LCQM.
We calculate the transition form factors and obtain the branching
ratios of the relevant decay modes in which a $\nu$, $\mu$ or $\tau$
lepton pair is produced at the order $10^{-6}\sim 10^{-7}$. For
$B_c\rightarrow D_s(1968)\ell \overline{\ell}$ decay modes, we give
a comparison of branching ratios with other predictions. The results
from our model are much larger than the results from RM and SM, and
are comparable with QM. We also give our predictions of branching
ratios of $B_c\rightarrow D_s^*(2317)\ell \overline{\ell}$ decay
modes, and notice that they are about 80 percent smaller than the
relevant ones of $B_c\rightarrow D_s(1968)\ell \overline{\ell}$
decay modes.

As the LHC started running recently, the $B_c$ meson plays an
important role in investigating the structure of hadrons and in
testing the unitarity of CKM quark mixing matrix. Experiments at the
LHC may not be able to measure the modes in which a neutrino pair is
produced. For $B_c\rightarrow D_s(1968)\ell^+\ell^-$, candidates for
$D_s(1968)$ can be reconstructed in the mode
$D_s\rightarrow\phi(\rightarrow K^+K^-)\pi$ \cite{CDF09}. To enhance
the search sensitivity, it can also be reconstructed in the modes
$D_s^+\rightarrow \overline{K}^{*0}K^+$, $D_s^+\rightarrow K^+
K_S^0$, or $D_s^+\rightarrow \pi^+\pi^+\pi^-$ \cite{CDF08,Belle09}.
If we take the mode $D_s\rightarrow \phi\pi$ of which the branching
ratio is $4.5\pm0.4\%$ into account to reconstruct $D_s$, the
effective branching ratio of $B_c\rightarrow D_s(\rightarrow
\phi\pi)\mu^+\mu^-$ is $0.56^{+0.48}_{-0.13}\times10^{-8}$ and that
of $B_c\rightarrow D_s(\rightarrow \phi\pi)\tau^+\tau^-$ is
$1.66^{+2.48}_{-1.08}\times10^{-9}$. For $B_c\rightarrow
D_s^*(2317)\ell^+\ell^-$, candidates for $D_s^*(2317)$ can be
reconstructed in the mode $D_s^*\rightarrow D_s\pi^0$
\cite{BABARetal1,BABARetal2,BABARetal3}. With the branching ratios
of the other decay modes of $D_s^*(2317)$ much smaller than that of
$D_s\pi^0$, the search sensitivity depends mostly on the
reconstruction of $D_s$. Therefore, the effective branching ratio of
$B_c\rightarrow D_s^*(\rightarrow D_s(\rightarrow
\phi\pi)\pi^0)\mu^+\mu^-$ is $1.02^{+0.86}_{-0.25}\times10^{-9}$ and
that of $B_c\rightarrow D_s^*(\rightarrow D_s(\rightarrow
\phi\pi)\pi^0)\tau^+\tau^-$ is $1.59^{+3.28}_{-0.97}\times10^{-10}$.
All the results predicted in this paper can be tested in the future
planned experiments at the LHC.

\section*{Acknowledgements}This work is partially supported by National Natural Science
Foundation of China (Grants No.~11021092, No.~10975003,
No.~11035003, and No.~11120101004).

\section*{Appendix A}
The form factors $f_+(q^2)$,~$F_T(q^2)$,~$u_+(q^2)$,~$U_T(q^2)$ can
be obtained directly from the calculation of matrix elements
$\langle D_s|\bar{s}\Gamma^{+}b|B_c\rangle$ in LCQM. They can be
expressed in explicit forms as:
\begin{eqnarray}
f_+(q^2)&=&\int_0^1\mathrm{d}x\int
\frac{\mathrm{d}^2\textbf{k}_{\perp}}{16{\pi}^3}\phi_s^*(x,\textbf{k}'_\perp)\phi_s(x,\textbf{k}_\perp)\nonumber\\
&\times&\frac{\mathcal{A}_s\mathcal{A}_b+\textbf{k}'_{\perp}\cdot
\textbf{k}_{\perp}}{\sqrt{\mathcal{A}_s^2+\textbf{k}'^2_{\perp}}\sqrt{\mathcal{A}_b^2+\textbf{k}^2_{\perp}}},\nonumber\\
u_+(q^2)&=&\int_0^1\mathrm{d}x\int
\frac{\mathrm{d}^2\textbf{k}_{\perp}}{16{\pi}^3}\frac{\widetilde{M_0}^2}{2\sqrt{3}M_0}\phi_p^*(x,\textbf{k}'_\perp)\phi_s(x,\textbf{k}_\perp)\nonumber\\
&\times&\frac{\mathcal{A'}_s\mathcal{A}_b+\textbf{k}'_{\perp}\cdot
\textbf{k}_{\perp}}{\sqrt{\mathcal{A}_s^2+\textbf{k}'^2_{\perp}}\sqrt{\mathcal{A}_b^2+\textbf{k}^2_{\perp}}},\nonumber\\
F_T(q^2)&=&\int_0^1\mathrm{d}x\int
\frac{\mathrm{d}^2\textbf{k}_{\perp}}{16{\pi}^3}\phi_s^*(x,\textbf{k}'_\perp)\phi_s(x,\textbf{k}_\perp)\nonumber\\
&\times&\frac{x(M_{B_c}+M_{D_s})\big[\mathcal{A}_b+(m_s-m_b)\frac{\textbf{k}_{\perp}\cdot
\textbf{q}_{\perp}}{\textbf{q}_{\perp}^2}\big]}{\sqrt{\mathcal{A}_s^2+\textbf{k}'^2_{\perp}}\sqrt{\mathcal{A}_b^2+\textbf{k}^2_{\perp}}},\nonumber\\
U_T(q^2)&=&\int_0^1\mathrm{d}x\int
\frac{\mathrm{d}^2\textbf{k}_{\perp}}{16{\pi}^3}\frac{\widetilde{M_0}^2}{2\sqrt{3}M_0}\phi_p^*(x,\textbf{k}'_\perp)\phi_s(x,\textbf{k}_\perp)\nonumber\\
&\times&\frac{x(M_{B_c}+M_{D_s^*})\big[\mathcal{A}_b-(m_s+m_b)\frac{\textbf{k}_{\perp}\cdot
\textbf{q}_{\perp}}{\textbf{q}_{\perp}^2}\big]}{\sqrt{\mathcal{A}_s^2+\textbf{k}'^2_{\perp}}\sqrt{\mathcal{A}_b^2+\textbf{k}^2_{\perp}}}, \nonumber\\
\end{eqnarray}
where $\textbf{k}'_{\perp}=\textbf{k}_{\perp}-x\textbf{q}_{\perp}$,
$\mathcal{A}_s=m_sx+m_q(1-x)$, $\mathcal{A}_b=m_bx+m_q(1-x)$, and
$\mathcal{A}_s=-m_sx+m_q(1-x)$.

For $f_-(q^2)$ and $u_-(q^2)$, we can not evaluate them by choosing
the plus component of the current, so we use the $\perp$ components
of the current to obtain $f_-(q^2)$ and $u_-(q^2)$:
\begin{align}
&\langle D_s|\bar{s}(\textbf{q}_{\perp}\cdot\gamma_{\perp})\gamma_5
b|B_c\rangle=\textbf{q}_{\perp}^2[f_+(q^2)-f_-(q^2)]\nonumber\\
=&\int
\frac{\mathrm{d}x\mathrm{d}^2\textbf{k}_{\perp}}{16{\pi}^3}\frac{x\phi_s^*(x,\textbf{k}'_\perp)\phi_s(x,\textbf{k}_\perp)}{\sqrt{\mathcal{A}_s^2+\textbf{k}'^2_{\perp}}\sqrt{\mathcal{A}_b^2+\textbf{k}^2_{\perp}}}\nonumber\\
\times&\big\{\frac{{\mathcal{A}_b^2+\textbf{k}^2_{\perp}}}{(1-x)x}(\textbf{k}_{\perp}+\textbf{q}_{\perp})\cdot\textbf{q}_{\perp}-\frac{{\mathcal{A}_s^2+\textbf{k}'^2_{\perp}}}{(1-x)x}\textbf{k}_{\perp}\cdot\textbf{q}_{\perp}\nonumber\\
+&[(-m_s+m_b)^2+\textbf{q}_{\perp}^2]\textbf{k}_{\perp}\cdot\textbf{q}_{\perp}\big\},
\end{align}
\begin{align}
&\langle
D_s^*|\bar{s}(\textbf{q}_{\perp}\cdot\gamma_{\perp})\gamma_5
b|B_c\rangle=\textbf{q}_{\perp}^2[u_+(q^2)-u_-(q^2)]\nonumber\\
=&\int
\frac{\mathrm{d}x\mathrm{d}^2\textbf{k}_{\perp}}{16{\pi}^3}\frac{\widetilde{M_0}^2}{2\sqrt{3}M_0}\frac{x\phi_p^*(x,\textbf{k}'_\perp)\phi_s(x,\textbf{k}_\perp)}{\sqrt{\mathcal{A}_s^2+\textbf{k}'^2_{\perp}}\sqrt{\mathcal{A}_b^2+\textbf{k}^2_{\perp}}}\nonumber\\
\times&\big\{\frac{{\mathcal{A}_b^2+\textbf{k}^2_{\perp}}}{(1-x)x}(\textbf{k}_{\perp}+\textbf{q}_{\perp})\cdot\textbf{q}_{\perp}-\frac{{\mathcal{A'}_s^2+\textbf{k}'^2_{\perp}}}{(1-x)x}\textbf{k}_{\perp}\cdot\textbf{q}_{\perp}\nonumber\\
+&[(m_s+m_b)^2+\textbf{q}_{\perp}^2]\textbf{k}_{\perp}\cdot\textbf{q}_{\perp}\big\}.
\end{align}

\def\Journal#1#2#3#4{{#1} {\bf#2}, #3 (#4)}
\def\PLB{{\rm Phys. Lett.}  B}
\def\PRT{\rm Phys. Rep.}
\def\PRL{\rm Phys. Rev. Lett.}
\def\PRD{{\rm Phys. Rev.} D}
\def\PRC{{\rm Phys. Rev.} C}
\def\ZPC{{\rm Z. Phys.} C}
\def\ZPA{{\rm Z. Phys.} A}
\def\AP{{\rm Ann. Phys.} }
\def\AM{{\rm Ann. Math.} }
\def\NP{{\rm Nucl. Phys.} B}
\def\RMP{\rm Rev. Mod .Phys}
\def\EPJ{{\rm Eur. Phys. J.} C}
\def\EPJA{{\rm Eur. Phys. J.} A}
\def\APP{{\rm Acta. Phys. Pol.} B}
\def\JPG{{\rm J. Phys.} G}
\def\CPC{\rm Comput. Phys. Commun}

\end{document}